\documentclass[prl,floatfix,twocolumn,showpacs,preprintnumbers,amsmath,amssymb,superscriptaddress]{revtex4}
\usepackage{graphicx,float,datetime}
\usepackage[ansinew]{inputenc}

\newdateformat{mydate}{\THEDAY\hspace{3pt}\monthname[\THEMONTH] \THEYEAR}

\begin{document}

\begin{center}
\title{Exact Static Cylindrical Black Hole Solution to Conformal Weyl Gravity}
\date{\mydate\today}
\author{Jackson Levi Said\footnote{jsai0004@um.edu.mt}}
\affiliation{Physics Department, University of Malta, Msida, MSD 2080, Malta}
\author{Joseph Sultana\footnote{joseph.sultana@um.edu.mt}}
\affiliation{Mathematics Department, University of Malta, Msida, MSD
2080, Malta}
\author{Kristian Zarb Adami\footnote{kristian.zarb-adami@um.edu.mt}}
\affiliation{Physics Department, University of Malta, Msida, MSD 2080, Malta}
\affiliation{Physics Department, University of Oxford, Oxford, OX1 3RH, United Kingdom}

\begin{abstract}
{We present the exact neutral black string solution in locally
conformal invariant Weyl gravity. As a special case, the general
relativity analogue still can be attained; however, only as a
sub-family of solutions. Our solution contains a linear term that
would thus result in a potential that grows linearly  over large
distances. This may have implications for exotic astrophysical
structures as well as matter fields on the extremely small scale.}
\end{abstract}


\maketitle

\end{center}

\section{I. Introduction}
There is a broad consensus in the community that Einstein's theory
of  general relativity describes well the behavior of gravitating
bodies in the solar system. This arose partly because of the
remarkable successes of general relativity in predicting the
outcomes of the classical tests. However, when the galactic scale and
beyond is considered, there has been a divide; in particular, the
split can be classified into two broad categories. One method
advanced is that of mass inference, where dark matter and then dark
energy is placed in the galactic and then the cosmological scale
such that the observations are an outcome of the theory using
standard Newtonian dynamics. The other, which has received much
attention in recent years, \cite{p00} is where the model of gravity
is modified in some way such that either no or little dark matter is
added. The proviso of these models is, however, that in the solar
system scale and up to current testability, the theories must agree
with general relativity in order to preserve the gains made thus
far.
\newline

The choice of the Einstein-Hilbert action, which produces general
relativity, is constrained by the requirement that the resulting
equations of motion be no higher than second order. This renders
field equations that are relatively simple compared with other
theories, but which fail to describe observations on scales much
higher than the solar system without placing a large amount of dark
matter to account for the resulting galactic rotational
curves. For this reason, a growing number of alternative actions are
being pursued, some driven by observation, others by foundational
development. One such model of the latter type is conformal
Weyl gravity, introduced in Refs.\cite{conformal1,conformal2,conformal3}.
This theory employs the principle of local conformal
invariance of the spacetime manifold as the  supplementary condition
that fixes the gravitational action, instead of the requirement that
the equations of motion be no higher than second order. This leads
to fourth order equations of motion for the gravitational field.
Nonetheless, the assumption of the local invariance principle,
besides being in line with the way actions are chosen in field
theory, leads to a unique action of conformal Weyl gravity among
all other fourth order theory actions. Furthermore, this invariance
principle may also provide a better link with the fundamental
quantum nature of reality due to the added symmetry inherent in the
model.
\newline

The outline of the paper is as follows. In Sec. II, we give
an introduction of the field equations of conformal Weyl
gravity with a brief review of the work done on the
spherically symmetric case. In Sec. III, we derive a
static and cylindrically symmetric solution to the field
equations in Weyl gravity and compare it with the static
cylindrically symmetric solution in general relativity obtained in
Ref.\cite{p01}. Some characteristics of this new solution,
such as the temperature and surface gravity of the cylindrical event
horizon, are obtained in Sec. IV. Finally, we end in Sec. V with a
discussion and some conclusions. Note that we use units where
$G=1=c$.

\section{II. Conformal Gravity}
The action for general relativity is given by
\begin{equation}
S=-\frac{1}{16\pi}\displaystyle\int d^4x\sqrt{-g}\left(R-2\Lambda\right),
\label{gr_action}
\end{equation}
where the action for the observer is suppressed, $g$ is the
determinant of the metric tensor, and $R$ is the Ricci scalar.
Despite passing the classical tests and every observation to date,
an issue naturally arises in the derivation of this action: Namely
why should the condition that the resulting equations of motion be
second-order guide the choice of action for gravity? That is, should
not the order cutoff be chosen in accordance with some postulate of
nature? One possible alternative, that when constrained
appropriately can still comply with the classical tests, is one
based on a local invariance principle. Besides the local gauge
invariance to which general relativity is subject, we consider the
singular restriction of local conformal invariance in choosing the
action for our gravity theory. This means that the manifold
$\left(\mathcal{M},\,g\right)$ which emerges must be invariant to
local stretchings \cite{p02}
\begin{equation}
g_{\mu\nu}\left(x\right)\rightarrow\tilde{g}_{\mu\nu}\left(x\right)=\Omega^2\left(x\right)\;g_{\mu\nu}\left(x\right),
\label{metr_conf_trans}
\end{equation}
where $x$ represents spacetime positions on the manifold, and where
the argument is not suppressed to emphasize that the conformal
invariance takes place locally.
\newline

One of the immediate consequences of this postulate is that the
artificially implanted cosmological constant, $\Lambda$, present in
the action for general relativity in Eq.(\ref{gr_action}), must be
withdrawn, since not to do so would introduce a length scale and thus
break the conformal symmetry in the theory. However, as will be shown
later, the same term naturally emerges out of the metric, which
provides further circumstantial evidence for the effectiveness of
the principle under consideration.
\newline

Turning now to the Weyl tensor
\begin{align}
C_{\lambda\mu\nu\kappa}&=R_{\lambda\mu\nu\kappa}\nonumber\\
&-\frac{1}{2}\left(g_{\lambda\nu}R_{\mu\kappa}-g_{\lambda\kappa}R_{\mu\nu}-g_{\mu\nu}R_{\lambda\kappa}+g_{\mu\kappa}R_{\lambda\nu}\right)\nonumber\\
&+\frac{1}{6}R\left(g_{\lambda\nu}g_{\mu\kappa}-g_{\lambda\kappa}g_{\mu\nu}\right),
\end{align}
which satisfies the conformal invariance condition \cite{p13}
\begin{equation}
C_{\lambda\mu\nu\kappa}\rightarrow\tilde{C}_{\lambda\mu\nu\kappa}=\Omega^2\left(x\right)\;C_{\lambda\mu\nu\kappa},
\end{equation}
necessary to render a theory in the first place. Hence, this can be
safely taken to be the unique Lagrangian density for Weyl gravity,
since due to its very locally conformal invariant nature, it must be
unique up to such transformations.
\newline

The consequence of this is that the Weyl action then becomes
\cite{conformal2}
\begin{align}
I_W&=\displaystyle\int d^4x\sqrt{-g} L\nonumber\\
&=-\alpha\displaystyle\int d^4x\sqrt{-g}C_{\lambda\mu\nu\kappa}C^{\lambda\mu\nu\kappa}\nonumber\\
&=-2\alpha\displaystyle\int d^4x\sqrt{-g}\left[R_{\mu\nu}R^{\mu\nu}-\frac{1}{3}R^2\right],
\label{weyl_action}
\end{align}
where the last equality is the simplest representation of the
action, obtained by applying the Lanczos identity \cite{p14} to the
Weyl tensor. Also, $\alpha$ is a dimensionless parameter which is
usually chosen to be positive in order to satisfy the Newtonian
lower limit. This parameter is the coupling constant of conformal
gravity.
\newline

It has become quite popular recently to take the Einstein-Hilbert
action present in Eq.(\ref{gr_action}) and add terms which then
vanish as the scale of the phenomena is reduced so that they take
place within the solar system. This method of generating extra
fields that make part of gravity comes in various forms, among which
are \cite{p04} $f\left(R\right)$ and \cite{p15} $f\left(T\right)$
gravity, where $T$ is the torsion scalar. However the distinction
between these additional terms and Weyl gravity is that the action
itself is different; that is, the driving force of change is the
additional local conformal invariance postulate in the theory. Thus,
instead of implanting a term in the action to explain new
phenomenology at higher scales, in Weyl gravity the conformal
constraint results as an outcome in the theory that only predicts
divergences from Einstein-Hilbert theory when scales greater than
the solar system are considered, closing at least one chapter in
general relativity that of the order of the theory.
\newline

Taking now the variation of the action in Eq.(\ref{weyl_action})
with respect to the metric leads to the field equations \cite{p02}
\begin{align}
\sqrt{-g}g_{\mu\alpha}g_{\nu\beta}\frac{\delta I_{W}}{\delta
g_{\alpha\beta}}
&=-2\alpha W_{\mu\nu}\nonumber\\
&=-\frac{1}{2}T_{\mu\nu}, \label{weyl_field_eqns}
\end{align}
where $T_{\mu\nu}$ is the stress-energy tensor and
\begin{equation}
W_{\mu\nu} = 2C^{\alpha\ \ \beta}_{\ \mu\nu\ ;\alpha\beta} +
C^{\alpha\ \ \beta}_{\ \mu\nu}R_{\alpha\beta}, \label{w1}
\end{equation}
is the Bach tensor. Incidentally, due to the form of the field
equations, whenever the Ricci tensor $R_{\mu\nu}$ vanishes,
$W_{\mu\nu}$ also vanishes so that every vacuum solution of
Einstein-Hilbert gravity also leads to a solution of Weyl gravity,
and thus all the work done carries on into Weyl gravity naturally
and without the need of alteration. Given, however, the increase in
complexity in Weyl gravity, the converse does not automatically
follow, meaning that not every vacuum solution from Weyl
gravity implies a solution for general relativity, so that a new
class of solutions are borne out.
\newline

This property emerges from the fact that $W_{\mu\nu}$ vanishes not
only when the Ricci tensor vanishes, but also by other means, which
can be seen by looking at Eqs.(\ref{weyl_field_eqns}) and
(\ref{w1}). Hence, the tensor that now characterizes the
geometry, $W_{\mu\nu}$, will not exactly replace the Ricci tensor.
Furthermore, in general, the fourth-order equations that make up
$W_{\mu\nu}$ contain a several-fold increase in complexity for the
field equations, which calls into question whether it will indeed be
possible to find a solution with more generality than those found in
general relativity.
\newline

For the task at hand and the problem addressed in this paper that
of determining the vacuum cylindrical spacetime line element we
consider the metric tensor in its most general form possible,
$g_{\mu\nu}=\mbox{diag}\left(-b\left(\rho\right),\,a\left(\rho\right),\,c\left(\rho\right),\,d\left(\rho\right)\right)$,
with the only further condition imposed that the spacetime is
static, since we are interested currently in the simpler case of a
neutral nonrotating spacetime.
\newline
At present, a number of conformal solutions have been found
\cite{conformal2,furthersolutions1,furthersolutions2,furthersolutions3,furthersolutions4,p03,p05,p06,p07}
and studied in quite some detail. The case of spherical symmetry was
studied by Mannheim and Kazanas in Ref.\cite{conformal2}, where the
initial problem seemed intractable but after a number of coordinate
transformations a solution was indeed found. This exact static and
spherically symmetric vacuum solution is given, up to a conformal
factor, by the metric
\begin{equation}
ds^2 = -B(r)dt^2 + \frac{dr^2}{B(r)} + r^2(d\theta^2 + \sin^2\theta
d\phi^2), \label{generalmetric}
\end{equation}
where
\begin{equation}
B(r) = 1 - \frac{\beta(2 - 3\beta\gamma)}{r} - 3\beta\gamma + \gamma
r - k r^2, \label{eqmetric}
\end{equation}
and $\beta,\,\gamma,$ and $k$ are integration constants. This solution
encompasses the Schwarzschild solution ($\gamma = k = 0$) and the
Schwarzschild-de Sitter solution ($\gamma = 0$) as special cases. In
this solution, the parameter $\gamma$ measures the departure of Weyl
theory from general relativity, and so for small enough $\gamma$,
both theories have similar predictions. This parameter has
dimensions of acceleration, and so Eq.(\ref{eqmetric}) provides a
characteristic, constant acceleration, which may be associated (in a
non-obvious way) with the cosmological setting. Given the
asymptotically non-flat character of the solution the parameter
$\gamma$ has been associated \cite{conformal2} with the inverse
Hubble length, i.e., $\gamma \simeq 1/R_H$, which for a typical
galaxy implies that the effects from the linear $\gamma r$ term in
the metric become comparable to those due to the Newtonian potential
term $2\beta/r$ on distance scales roughly equal to the size of the
galaxy a fact that led Mannheim and Kazanas to produce fits to the
galactic rotation curves. The effect of the linear term $\gamma r$
in the metric on classical tests, particularly the bending of light,
has been studied in detail in
Refs.\cite{deflection1,deflection2,deflection3,deflection4,deflection5}
The solution Eq.(\ref{eqmetric}) was also generalized both to rotating
and charged solutions in Ref.\cite{furthersolutions1}, heralding the
complete generalization of spherical symmetry from general
relativity into Weyl gravity.
\newline

Following the success of spherical symmetry in conformal
gravity, research shifted to topological black holes, which
culminated in Refs.\cite{p03,p05}, where the question of conformal
topological black holes was explored in general terms. Besides
providing new solutions, these works showed that in conformal gravity,
topological black hole solutions with non-negative scalar curvature
$k$ at infinity are possible, unlike general relativity where only
asymptotically anti-de Sitter (AdS) topological black holes are
possible. The only exception is the toroidal case,
$\mathcal{S}^1\times\mathcal{S}^1$, where the black hole
interpretation is only possible for $k < 0$ as in AdS gravity.
\newline

Besides compact spacetimes, cosmological effects have also been
studied. Indeed, in Ref.\cite{p12} cosmological conformal gravity
fluctuations were studied and the dark energy problem is discussed
in this setting. In particular, the theory is put in a different
setting so that it can be set against cosmological data, thus
providing the way forward for further study into the local
divergences from the isotropic and homogenous cosmos.
\newline

We now focus on the class of spacetimes in Weyl gravity with
cylindrical symmetry. A number of analytic and numerical solutions
were presented in Refs.\cite{p06,p07}, including a generalization of
the Melvin solution, as well as a study of the magnetic properties
of conformal cylindrical solutions. However, the cylindrical
solutions were obtained in a gauge that did not naturally generalize
the well-known black string  solution in general relativity, given
by the Lemos metric \cite{p01}
\begin{align}
ds^2&=-\left(\alpha^2r^2-\frac{b}{\alpha r}\right)dt^2+\frac{dr^2}{\alpha^2r^2-\frac{b}{\alpha r}}\nonumber\\
&+r^2\,d\phi^2 + \alpha^2r^2\,dz^2, \label{lemos_metr}
\end{align}
with the coordinate ranges
\begin{align}
&-\infty<t<\infty,\,\, 0\leq r<\infty,\,\, 0\leq\phi<2\pi,\nonumber\\
& -\infty<z<\infty,
\end{align}
and with $\alpha=\sqrt{-\frac{\Lambda}{3}}$, $b=M/2$, and $M$ being mass.
\newline

Our aim is to find a Lemos-like black string solution in
Weyl gravity similar to what Mannheim and Kazanas did in the
spherically symmetric case discussed above, and then compare this
with the Lemos metric to study further any similarities and
differences between Weyl and Einstein's theories.

\section{III. The Conformal Cylindrical Metric}
In order to solve the conformal field equations with a cylindrically
symmetric metric tensor, we consider a general line element in
cylindrical coordinates $\left(t,\,\rho,\,\phi,\,z\right)$:
\begin{equation}
ds^2=-b\left(\rho\right)\,dt^2+a\left(\rho\right)\,d\rho^2+c\left(\rho\right)\,d\phi^2+d\left(\rho\right)\,dz^2.
\end{equation}

Since in cylindrical topology the background spacetime is not curved
along the axial direction or over the angular coordinate, the metric
elements will be independent of both $z$ and $\phi$. Moreover, since
we are looking for a conformal generalization of the Lemos metric,
Eq.(\ref{lemos_metr}), in the spirit of Ref.\cite{conformal2}, we
can consider the Lemos gauge such that
\begin{equation}
C(\rho) = \rho^2\quad\mbox{and}\quad
D\left(\rho\right)=\alpha^2\rho^2.
\end{equation}
\newline

Following a similar procedure as in Ref.\cite{conformal2}, we
reformulate the metric in order to reduce the computation required
to solve the field equations. The line element may be rewritten as
\begin{align}
ds^2&=\frac{\rho^2\left(r\right)}{r^2}\big[-\frac{r^2 b\left(r\right)}{\rho^2\left(r\right)}\,dt^2+\frac{r^2 a\left(r\right)\rho'^{2}(r)}{\rho^2\left(r\right)}\,dr^2\nonumber\\
&+r^2\,d\phi^2+\alpha^2 r^2\,dz^2\big],
\end{align}
where $\rho\left(r\right)$ is an arbitrary function of $r$. Choosing
this dependence such that
\begin{equation}
\displaystyle\int\frac{d\rho}{\rho^2\left(r\right)}=-\frac{1}{\rho\left(r\right)}=\displaystyle\int\frac{dr}{r^2\sqrt{a\left(r\right)b\left(r\right)}},
\end{equation}
then yields
\begin{align}
ds^2&=\frac{\rho^2(r)}{r^2}\bigg[-B\left(r\right)\,dt^2+A\left(r\right)\,dr^2+r^2\,d\phi^2\nonumber\\
&+\alpha^2 r^2\,dz^2\bigg],
\end{align}
with $A\left(r\right)=1/B(r)$ and $B\left(r\right)=\frac{r^2
b\left(r\right)}{\rho^2\left(r\right)}$. The metric that results is
thus conformally related to the standard general line element for
cylindrical spacetimes with $A(r) = 1/B(r)$. Conformal
transformations are allowed through Eq.(\ref{metr_conf_trans}), so we
take
\begin{equation}
g_{\mu\nu}\rightarrow r^2\rho^{-2}\left(r\right)g_{\mu\nu},
\end{equation}
and hence the general line element
\begin{equation}
ds^2=-B\left(r\right)\,dt^2+\frac{dr^2}{B\left(r\right)}+r^2\,d\phi^2+\alpha^2
r^2\,dz^2, \label{con_cyl_met}
\end{equation}
is formulated. Hence, the metric elements may now be determined up to
an arbitrary overall $r$-dependent conformal factor. Furthermore,
since vacuum solutions of $W_{\mu\nu}\left(\rho\right)$ will be
considered, $W_{\mu\nu}\left(r\right)$ must also vanish, from which
it follows that the information is completely transferred to
$W_{\mu\nu}\left(r\right)$ so that this will contain all observable
information in the vacuum case.
\newline

The method used in general relativity of calculating the Ricci
tensors and then equating them to a vanishing stress-energy tensor
becomes far too complicated in conformal gravity given
Eq.(\ref{w1}). Indeed, this case strengthens when the covariant
derivative is considered. Instead, as was done in Ref.\cite{conformal2},
we calculate the Euler-Lagrange equations using the generic line
element
\begin{equation}
ds^2=-B\left(r\right)\,dt^2+A\left(r\right)\,dr^2+r^2d\,\phi^2+\alpha^2r^2\,dz^2,
\label{metr_drv_init}
\end{equation}
which adopts the Lemos gauge for cylindrically symmetric spacetimes,
and then substitute $A\left(r\right)=1/B\left(r\right)$.
\newline

The Euler-Lagrange equations turn out to be second order
\cite{conformal2}
\begin{align}
&\sqrt{-g}\,W^{\mu\mu}=\frac{\delta I}{\delta g_{\mu\mu}}=\frac{\partial}{\partial g_{\mu\mu}}\left(\sqrt{-g}\tilde{L}\right)\nonumber\\
&-\frac{\partial}{\partial x^{\mu}}\left(\sqrt{-g}\frac{\partial
\tilde{L}}{\partial
\left(g_{\mu\mu}\right)'}\right)+\frac{\partial^2}{\partial
\left(x^{\mu}\right)^2}\left(\sqrt{-g}\frac{\partial
\tilde{L}}{\partial \left(g_{\mu\mu}\right)''}\right), \label{el}
\end{align}
where ' indicates differentiation with respect to $r$ and $\tilde{L}
= R_{\mu\nu}R^{\mu\nu} - R^2/3$. We consider only the diagonal
elements of $W^{\mu\nu}$ which only vary with respect to the radial
coordinate since the line element in Eq.(\ref{metr_drv_init}) has
elements that depend only on that coordinate. The previous variation is
taken for $\delta I/\delta A$ and $\delta I/\delta B$, respectively,
yielding
\begin{widetext}
\begin{align}
\sqrt{\alpha^2r^4AB}\,W^{rr}&= -\frac{\alpha ^2}{48A(r)^4 B(r)^3\sqrt{r^4\alpha^2A\left(r\right)B\left(r\right)}}\bigg[-7 r^2 B(r)^2 A'(r)^2 \left(r B'(r)-2 B(r)\right)^2\nonumber\\
&+2 r^2 A(r) B(r) \left(2 B(r)-r B'(r)\right) \big(4B(r)^2 A''(r)+3 r A'(r) B'(r)^2\nonumber\\
&-2 B(r) \left(r A''(r) B'(r)+A'(r) \left(2 r B''(r)+B'(r)\right)\right)\big)\nonumber\\
&+A(r)^2\bigg(-7 r^4 B'(r)^4+4 r^3 B(r) B'(r)^2 \left(3 r B''(r)+5 B'(r)\right)\nonumber\\
&+4 r^2 B(r)^2 \left(r^2 B''(r)^2+B'(r)^2-2 r B'(r)\left(r B^{(3)}(r)+6 B''(r)\right)\right)\nonumber\\
&+16 r B(r)^3 \left(r \left(r B^{(3)}(r)+2 B''(r)\right)-2 B'(r)\right)+16B(r)^4\bigg)\bigg],
\label{wrr}
\end{align}
\end{widetext}
and
\begin{widetext}
\begin{align}
\sqrt{\alpha^2r^4AB}\,W^{tt}&=\frac{-\alpha ^2}{48A(r)^4B(r)^4\sqrt{r^4\alpha^2A(r)B(r)}}\bigg[ 56 r^3 B(r)^3 A'(r)^3 \left(r B'(r)-2 B(r)\right)\nonumber\\
&+r^2 A(r) B(r)^2 A'(r) \bigg(57 r^2 A'(r) B'(r)^2-4 r B(r)\bigg(13 r A''(r) B'(r)\nonumber\\
&+A'(r) \left(19 r B''(r)+13 B'(r)\right)\bigg)+4 B(r)^2 \left(26 r A''(r)+7 A'(r)\right)\bigg)\nonumber\\
&+2r A(r)^2 B(r) \bigg(29 r^3 A'(r) B'(r)^3-6 r^2 B(r) B'(r) \big(2 r A''(r) B'(r)\nonumber\\
&+A'(r) \left(9 r B''(r)+4B'(r)\right)\big)+4 r B(r)^2 \bigg(A'(r) \left(r \left(6 r B^{(3)}(r)+13 B''(r)\right)-5 B'(r)\right)\nonumber\\
&+r \left(4 r A''(r)B''(r)+\left(r A^{(3)}(r)+3 A''(r)\right) B'(r)\right)\bigg)\nonumber\\
&+8 B(r)^3 \left(2 A'(r)-r \left(rA^{(3)}(r)+A''(r)\right)\right)\bigg)+A(r)^3 \bigg(-16 r^3 \left(4 B^{(3)}(r)+r B^{(4)}(r)\right) B(r)^3\nonumber\\
&+49 r^4 B'(r)^4-4r^3 B(r) B'(r)^2 \left(29 r B''(r)+11 B'(r)\right)\nonumber\\
&+4 r^2 B(r)^2 \left(9 r^2 B''(r)^2-5 B'(r)^2+2 r B'(r) \left(6 rB^{(3)}(r)+13 B''(r)\right)\right)+16 B(r)^4\bigg)\bigg].
   \label{wtt}
\end{align}
\end{widetext}
The other two elements, $W^{\phi\phi}$ and $W^{zz}$, do not need to be
taken into account, since we have a sufficient number of constraints.
These two further equations provide us with an independent check of
any solution that results.
\newline

We now restrict ourselves to the line element in
Eq.(\ref{metr_drv_init}), taking $A\left(r\right)$ to be the
reciprocal of $B\left(r\right)$. This implies that we need only one
constraint to determine the line element, since $B\left(r\right)$
turns out to be the only unknown. Hence, all the information is
contained in $W^{rr}$, irrespective of the number of derivatives it
contains. Taking the vacuum solution, $W^{rr}=0$, the resulting
equation to solve becomes
\begin{align}
&r^2 \bigg(-r^2 B''(r)^2-4 B'(r)^2\nonumber\\
&+2 r B'(r) \left(r B^{(3)}(r)+2 B''(r)\right)\bigg)\nonumber\\
&-4 r B(r) \left(r \left(rB^{(3)}(r)+B''(r)\right)-2 B'(r)\right)\nonumber\\
&-4 B(r)^2=0,
\label{step1}
\end{align}
which can in principle be solved analytically.
\newline

In order to reduce the overall order of the differential equation, we
transform $B\left(r\right)$ by
\begin{equation}
B\left(r\right)=r^2l\left(r\right),
\end{equation}
which enables us to rewrite Eq.(\ref{step1}) as
\begin{align}
&-r^2 l''(r)^2+8 l'(r)^2+2 r l'(r) \left(r l^{(3)}(r)+4 l''(r)\right)=0.
\label{step_2}
\end{align}
hence reducing the order when the substitution
\begin{equation}
l'\left(r\right)=y\left(r\right),
\end{equation}
is taken. However since, ignoring derivatives, the function
$l\left(r\right)$ appears twice in every term, we transform the
plane by the exponential function
\begin{equation}
y\left(r\right)=e^{f\left(r\right)},
\end{equation}
which surprisingly yields the relatively simple expression
\begin{align}
2 r^2 f''(r)+r^2 f'(r)^2+8 r f'(r)+8=0.
\end{align}
Again reducing the overall order of the differential equation through
\begin{equation}
f'\left(r\right)=h\left(r\right),
\end{equation}
so that the arbitrary function is constrained by
\begin{equation}
2 r^2 h'(r)+r^2 h(r)^2+8 r h(r)+8=0,
\end{equation}
the first integral is then
\begin{equation}
h\left(r\right)=\frac{1}{a+\frac{r}{2}}-\frac{4}{r},
\end{equation}
and repeating each substitution in reverse we arrive at the final
unknown in the line element in Eq.(\ref{metr_drv_init}):
\begin{equation}
B\left(r\right)=\frac{4 a^2 c}{2r}+2ac+cr+dr^2,
\label{solution}
\end{equation}
where $a$, $c$, and $d$ are constants of integration.
\newline

The result in Eq.(\ref{solution}) has all the terms of the Lemos
analogue; however, due to the placement of the constants, the exact
form of the Lemos line element in Eq.(\ref{lemos_metr}) cannot be
recovered. For this, we take the transformation
\begin{align}
a&=\sqrt{\frac{b_1^2b_2b_3-3b_4}{b_2b_3}},\\
c&=\frac{b_3}{4},\\
d&=b_2^2,
\end{align}
thus solving the problem of a conformal generalization of the Lemos
metric where $B\left(r\right)$ is
\begin{align}
B\left(r\right)&=\frac{b_1^2b_2b_3-3b_4}{3b_2r}+\sqrt{\frac{b_3\left(b_1^2b_2b_3-3b_4\right)}{4b_2}}\nonumber\\
&+\frac{b_3r}{4}+b_2^2r^2,
\end{align}
or more conveniently, in the form
\begin{equation}
A^{-1}\left(r\right)=B\left(r\right)=\frac{\beta}{r}+\sqrt{\frac{3\beta\gamma}{4}}+\frac{\gamma r}{4}+k^2r^2,
\label{f_solution}
\end{equation}
where
\begin{align}
\beta&=\frac{b_1^2b_2b_3-3b_4}{3b_2}\\
\gamma&=b_3\\
k&=b_2,
\end{align}
from which we can regain the Lemos line element by setting the
emergent conformal factor $\gamma$ to zero. Hence, the conformal
Lemos-like metric can be found without taking any approximations at
all. Moreover, the solution in Eq.(\ref{f_solution}) or Eq.(\ref{solution})
also satisfies the remaining vacuum field equations $W^{ii} = 0$ for
$i=t,\ \phi,\mbox{and}\ z$.
\newline

Given the metric components in Eq.(\ref{f_solution}), we compare the
conformal line element with the same line element derived in general
relativity in Eq.(\ref{lemos_metr}). This is achieved by setting the
new constant conformal factor $\gamma$ to zero and relating the
remaining components, which would imply that
\begin{align}
k&=\alpha,\\
\beta&=-\frac{b}{\alpha},
\end{align}
hence we recover the expected metric for general relativity \cite{p01}.
\newline

Lastly, we close by considering the curvature of the cylindrical
metric; in particular, the Ricci curvature invariant turns out to be
given by
\begin{equation}
\mathcal{R}=g^{\mu\nu}R_{\mu\nu}=-\frac{24k^2r^2+3r\gamma+2\sqrt{3\beta\gamma}}{2r^2}.
\end{equation}
The simplicity of this expression stems from the fact that the only
nonvanishing Ricci tensor components are on the diagonal and depend
only on one of the spacelike coordinates simplifying many of the
derivatives and sums. In a similar, way the Kretschmann scalar
invariant that results also turns out to be remarkably simple:
\begin{align}
&\mathcal{K}=R^{\lambda\mu\nu\sigma}R_{\lambda\mu\nu\sigma}=\frac{1}{2r^6}\Bigg[24\beta^2 + 6\beta\gamma r^2+\gamma^2 r^4 + \nonumber\\
&12\gamma k^2 r^5 + 48k^4 r^6 +8\sqrt{3\beta\gamma}r\left(\beta +
\frac{\gamma r^2}{4} + k^2 r^3\right)\Bigg].
\end{align}
Through the above scalar invariant, the physical singularity is found
to be located at $r=0$.
\newline

\section{IV. Temperature}
The Hawking temperature of the metric in Eq.(\ref{metr_drv_init})
with $A\left(r\right)$ and $B\left(r\right)$ as defined in
Eq.(\ref{f_solution}) may provide interesting insight into the
quantum nature of the surrounding spacetime. The surface gravity
that forms the underpinning of temperature by the relationship
$T_H=\frac{\kappa}{2\pi}$ is defined by the formula \cite{p16}
\begin{equation}
\kappa^2=-\frac{1}{2}\left(\nabla^{\mu}\chi^{\nu}\right)\left(\nabla_{\text{[}\mu}\chi_{\nu\text{]}}\right),
\label{sg}
\end{equation}
where $\chi^{\nu}$ is the Killing field generating the
horizon $r_{h}$ where $B(r_{h}) = 0$. This is given by
\begin{equation}
r_h=\frac{1}{12k^2}\left[-\gamma+\frac{\gamma^2-24k^2\sqrt{3b\gamma}}{\Sigma}+\Sigma\right],
\end{equation}
where
\begin{align}
\Sigma^3&=-864\beta k^4-\gamma^3+36\gamma k^2\sqrt{3\beta\gamma}+\nonumber\\
&12\sqrt{3\beta k^4\left(1728\beta k^4+\gamma^3-48\gamma
k^2\sqrt{3\beta\gamma}\right)},
\end{align}
In the case of metric Eq.(\ref{metr_drv_init}), the surface gravity
given by Eq.(\ref{sg}) gives
\begin{equation}
\kappa=-\frac{\beta}{2r^2}+\frac{\gamma}{8}+k^2r,
\end{equation}
and so the temperature is found to be
\begin{equation}
T_H=\frac{\kappa}{2\pi}=\frac{1}{\pi}\left(-\frac{\beta}{4r_h^2}+\frac{\gamma}{16}+\frac{k^2r_h}{2}\right).
\end{equation}
For $\gamma=0$ this reduces to the general relativity result \cite{p01} as expected.

\section{V. Conclusion}
The first point to note about our solution is that it is a vacuum
solution ($W_{\mu\nu} =0$) with no cosmological constant term in the
field equations, whereas the Lemos metric, which is retrieved by
taking $\gamma=0$ in Eq.(\ref{f_solution}), is a solution of
Einstein's field equations in AdS gravity. The constant $\gamma$
thus measures the departure of Weyl gravity from general relativity.
Comparing Eq.(\ref{f_solution}) with the Lemos metric in Eq.
(\ref{lemos_metr}), which is also a vacuum solution of conformal
gravity, we note that $\beta$ must be related to the negative of the
mass and $k = \alpha = \sqrt{-\Lambda/3}$ is related to the
cosmological constant.
\newline

General relativity is by design a strong field theory in that in the
weak field, it asymptotes to Newtonian gravity. Conformal (Weyl)
gravity, on the other hand, aims to derive a theory of gravity using
no restrictions on the order of the field equations. Applying just
the invariance principle in Eq. (\ref{metr_conf_trans}), one can
still achieve the standard Newtonian phenomenology in the weak field
limit, \cite{furthersolutions4} and possibly a solution to the dark
matter and dark energy problems.  The solution found in this paper
will not be applicable to most astrophysical sources, since for the
most part they are organized with spherical symmetry. On the other,
hand for theories such as string theory, the new solution presented
in this paper may have uses on scales just above the Planck scale
\cite{p27}. However, in any case, detection of any such objects would
in all likelihood be in the form of Hawking radiation, which is one
possible avenue of future development for this metric. There may
also be string-like applications in a number of other theories, such
as with the use of the AdS/CFT correspondence duality.
\newline

In this paper, we derive the static cylindrically neutral metric for
Weyl conformal gravity. The next step would be to generalize this to
the whole family of charged rotating cylindrically symmetric
solutions, as was done by Mannheim and Kazanas
\cite{furthersolutions1} for the Kerr-Neuman general relativity
analogue. This adds to the current cylindrically symmetric solutions
in Refs.\cite{p06,p07} in that the field equations are solved
analytically in a gauge which naturally generalizes the black string
solutions of Einstein's gravity.

\section{Acknowledgments}
J. L. S. wishes to thank the Physics Department at the University of
Malta for hospitality during the completion of this work.

\end{document}